\documentclass[aps,prl,twocolumn,showpacs,superscriptaddress,floatfix]{revtex4-2}
\usepackage{graphicx}
\usepackage{dcolumn}
\usepackage{bm}
\usepackage{hyperref}
\usepackage[utf8]{inputenc}
\usepackage[english]{babel}
\usepackage{units}
\usepackage{mathtools}
\hypersetup{colorlinks=true,citecolor={blue},linkcolor={blue},urlcolor={blue}}
\usepackage{float}
\usepackage{amsmath}
\usepackage{empheq}
\usepackage{mathrsfs}
\usepackage{amssymb}
\usepackage{soul}
\usepackage{textcomp}
\usepackage{placeins}
\usepackage{xcolor}

\setstcolor{red}

\begin{document}

\title{Engineering photon statistics in a spinor polariton condensate}

\author{S. Baryshev}
\address{Skolkovo Institute of Science and Technology, Moscow, Territory of innovation center “Skolkovo”,
Bolshoy Boulevard 30, bld. 1, 121205, Russia.}

\author{A. Zasedatelev}
\email{A.Zasedatelev@skoltech.ru}
\address{School of Physics and Astronomy, University of Southampton,  Southampton, SO17 1BJ, UK.}
\address{Skolkovo Institute of Science and Technology, Moscow, Territory of innovation center “Skolkovo”,
Bolshoy Boulevard 30, bld. 1, 121205, Russia.}

\author{H. Sigurdsson}
\address{School of Physics and Astronomy, University of Southampton,  Southampton, SO17 1BJ, UK.}
\address{Skolkovo Institute of Science and Technology, Moscow, Territory of innovation center “Skolkovo”,
Bolshoy Boulevard 30, bld. 1, 121205, Russia.}

\author{I. Gnusov}
\address{Skolkovo Institute of Science and Technology, Moscow, Territory of innovation center “Skolkovo”,
Bolshoy Boulevard 30, bld. 1, 121205, Russia.}

\author{J. D. Töpfer}
\address{School of Physics and Astronomy, University of Southampton,  Southampton, SO17 1BJ, UK.}

\author{A. Askitopoulos}
\address{Skolkovo Institute of Science and Technology, Moscow, Territory of innovation center “Skolkovo”,
Bolshoy Boulevard 30, bld. 1, 121205, Russia.}

\author{P. G. Lagoudakis}
\address{Skolkovo Institute of Science and Technology, Moscow, Territory of innovation center “Skolkovo”,
Bolshoy Boulevard 30, bld. 1, 121205, Russia.}
\address{School of Physics and Astronomy, University of Southampton,  Southampton, SO17 1BJ, UK.}

\begin{abstract}

We implement full polarization tomography on the photon correlations in a spinor exciton-polariton condensate. Our measurements reveal condensate pseudospin mean-field dynamics spanning from stochastic switching between linear polarization components, limit cycles, and stable fixed points, and their intrinsic relation to the condensate photon statistics. We optically harness the cavity birefringence, polariton interactions, and the optical orientation of the photoexcited exciton background to engineer photon statistics with precise control. Our results demonstrate a smooth transition from a highly coherent to a super-thermal state of the condensate polarization components. 

\end{abstract}
\maketitle
Photon statistics is of central importance in laser physics and quantum optics, and serves as an essential toolbox for the characterization of optical sources ranging from pure single-photon guns to super-thermal highly fluctuating light sources. In particular, the width of the photon distribution in a laser defines its noise properties, whose understanding is at the heart of many applications such as laser cooling ~\cite{Safavi_Naeini_2013}, precise interferometry~\cite{PhysRevD.23.1693}, optical communication~\cite{Banaszek:20} etc. Coherent shot-noise-limited light sources are a revolutionary resource in modern science and technology. While an ideal laser obeys Poisson photon distribution, practical devices usually suffer from excessive noise that broadens the distribution and affects phase stability. Mode competition is one  detrimental effect generating excessive, so-called, super-Poisson noise in conventional semiconductor microcavity lasers~\cite{PhysRevB.56.R7076,PhysRevLett.84.4337,PhysRevA.87.053819,Redlich_2016}. On the other hand, stochastic mode switching allows studying intriguing phenomena of chaos in photonic systems~\cite{chaos} and opens the door for new optoelectronic applications such as ghost imaging~\cite{ghost} and multiphoton microscopy~\cite{multyphotonmicro} with super-bunched light, while mode beating enables low-energy ultrafast optical communication~\cite{ultrafastspinlasers} etc. The intrinsic linear mode-coupling  usually dominates over nonlinear effects in conventional microlasers. However, in semiconductor structures with strong-light matter interactions, this may not be the case.

Exciton-polaritons (here on polaritons) are bosonic quasiparticles originating from strong light-matter coupling of excitons with photons in semiconductor microcavities~\cite{PhysRevLett.69.3314}. They can undergo a power-driven Bose–Einstein type condensation~\cite{Kasprzak} into a spinor order parameter $\Psi = (\psi_+,\psi_-)^T$ corresponding to the right-hand $\sigma^+$ and left-hand $\sigma^-$ circular polarization of the emitted light like in a conventional semiconductor spin-laser~\cite{SpincontVCSEL}.
Besides its fundamental importance, spin degree of freedom is particularly attractive for applications~\cite{polaritonicdevices}. The exciton component makes the polariton Bose gas inherently nonlinear due to pair-particle interactions, permitting experimental observation of quantum correlations~\cite{blockade, blockade2} and superfluidity~\cite{superfl}. Polariton condensate offers unprecedented all-optical control with a promise for efficient devices that range from amplifiers~\cite{PhysRevLett.84.1547,High-temperatureamplification}, transistors~\cite{poltrans}, tunnelling diodes~\cite{PhysRevLett.110.236601}, routers~\cite{doi:10.1063/1.4936158}, phase-controlled interferometers~\cite{phasemodint}, ultrafast switches~\cite{spinswitches,PhysRevB.85.235102} to volatile memory elements~\cite{Cerna_NatComm2013}. A dark side of polariton interactions is its detrimental role on coherence properties of the condensate. Coherence time of the condensate in most experiments is limited to $\sim10-100$ ps~\cite{Nardin_PRL2009,  PhysRevLett.109.016404, PhysRevX.6.011026} predominantly due to interactions with an incoherent photoexcited background exciton reservoir. Moreover, a noisy reservoir induces fast spin decoherence~\cite{PhysRevX.5.031002} which was evidenced through dephasing of the condensate polarization in second-order photon correlation measurements~\cite{PhysRevB.93.115313}. Until now, the experimental study of the condensate spinor mean-field dynamics in connection with photon statistics is barely investigated. In the work by Sala et al. it was demonstrated how stochastic spinor dynamics was affecting polarization-resolved photon statistics of the condensate~\cite{PhysRevB.93.115313}. There, pulsed excitation procedures and strong dephasing due to the presence of an exciton reservoir both obstructed study of the spinor dynamics and thus limits possible applications. 

In this work, we exploit an optical trap configuration to separate the polariton condensate from the exciton reservoir extending its coherence time over two orders of magnitude $\sim1$ ns compared to the polariton lifetime ~\cite{askitopoulos2019giant}. With such long coherence times we are able to use continuous wave non-resonant optical excitation and drive the spinor condensate into stable solutions described by its coherent mean field dynamics. In tandem with polarization resolved photon correlation tomography and precise spectroscopy we unravel complex spinor condensate dynamics and its connection to photon statistics. Importantly, we experimentally demonstrate two limiting cases of spinor dynamics governed by intrinsic energy splitting of the structure and nonlinear splitting induced by polariton interactions. 
We use our findings to engineer photon statistics of polariton condensates and demonstrate crossover from super-thermal photon distribution to a highly coherent state.

We investigate the photon statistics  between different pseudospin projections of the polariton condensate defined as,
\begin{align}\label{eq.1}
\mathbf{S} = \frac{1}{2} \Psi^\dagger \boldsymbol{\sigma} \Psi, 
\end{align} 
with $\boldsymbol{\sigma}$ being the Pauli matrix vector.  The condensate is both optically stimulated and trapped in a ring-shaped potential~\cite{Askitopoulos_PRB2013} under continuous wave (CW) non-resonant $\lambda_\text{ext}=783$~nm annular optical excitation of a GaAs/AlAs$_{0.98}$P$_{0.02}$ 2$\lambda$-microcavity with embedded InGaAs quantum wells~\cite{doi:10.1063/1.4901814} held in a cryostat at 4K. The linearly polarized laser beam is shaped by a spatial light modulator  such that it forms a $12~\mu$m diameter annular pump profile on the sample plane [Fig.~\ref{fig.1}, inset]. The beam is chopped by an acousto-optical modulator following a square waveform with 1~kHz frequency and 10\% duty cycle to avoid sample heating. We implement a polarization-resolved multi-channel Hanbury Brown and Twiss (HBT) intensity interferometer to measure photon correlations between the pseudospin projections $\mathbf{S} = (S_1,S_2,S_3)$ which correspond explicitly to the Stokes parameters of linear vertical-horizontal ($S_1$), diagonal-antidiagonal ($S_2$), and right-hand and left-hand circular ($S_3$) polarization of the emitted cavity light. We extract statistical information of the correspondent photon distributions from Stokes component second order auto-correlation measurements at HBT1 and HBT2 (e.g., vertical-vertical and horizontal-horizontal correlations). Moreover, our experimental setup allows simultaneous measurements of cross-correlations between orthogonal polarizations by means of a third interferometer HBT3 as shown in Fig.~\ref{fig.1} (e.g., vertical-horizontal correlations). 

\begin{figure}[hbt]
\includegraphics [width=1\columnwidth]{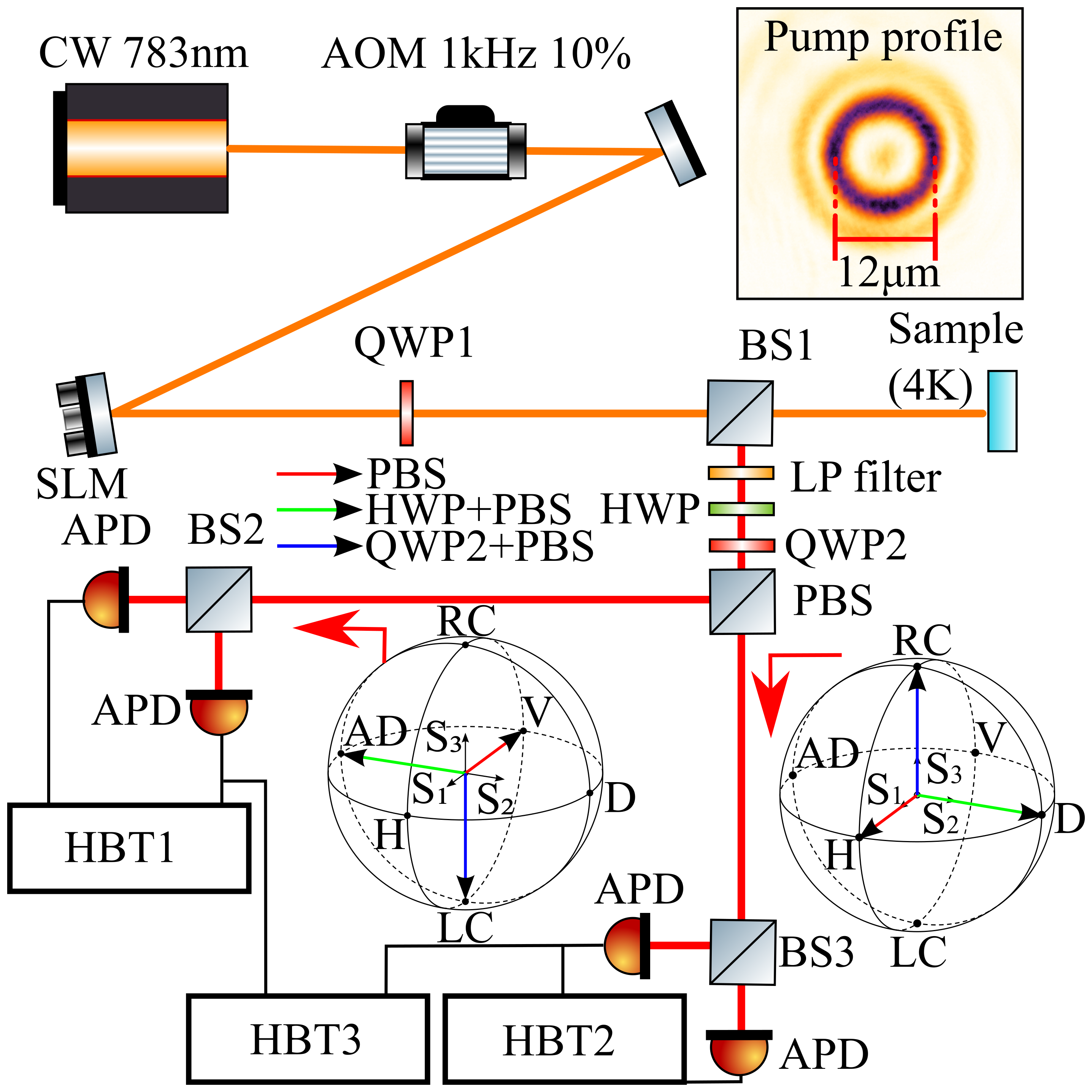}
\centering
\caption{Experimental setup for full polarization tomography of photon correlations for an optically trapped polariton condensate. AOM - acousto-optical modulator, SLM - spacial light modulator, QWP - quarter waveplate, HWP - half waveplate, BS - 50:50 beam splitter, PBS - polarizing beam splitter, LP - long pass filter, APD - avalanche photodetector. Inset shows the transverse laser intensity profile where the condensate forms in its center.}
\label{fig.1}
\end{figure}

The PBS, HWP and QWP2 components at the interferometer part of the setup allow us to realize a full correlation tomography of the orthogonal pseudospin projections on a Poincar\'e sphere in a time domain as schematically illustrate in Fig.~\ref{fig.1}. Time-correlated single photon counting technique~\cite{o2012time} in tandem with fast silicon avalanche photodetectors enable correlation measurements with $\tau = 60$~ps time resolution. Thus, the interferometer enables direct measure of second order photon auto- and cross-correlation functions $g_{i,j}^{(2)}$, where $i,j$ denote - horizontal (H), vertical (V); diagonal (D), antidiagonal (AD); and left-circular (LC), right-circular (RC) polarizations - in accordance with the well-known expression below for the second order-correlation function.
\begin{align}\label{eq.2}
g_{i,j}^{(2)}(\tau)= \frac{\langle{a_{i}^{\dag}(t)}{a_{j}^{\dag}(t+\tau)}{a_{j}(t+\tau)}{a_{i}(t)}\rangle}{\langle{a_{i}^{\dag}(t)}{a_{i}(t)}\rangle\langle{a_{j}^{\dag}(t+\tau)}{a_{j}(t+\tau)}\rangle},
\end{align}
where $a_{i}^\dagger$ and $a_{i}$ are photon creation and annihilation operators for given polarizations $i,j$ and $\tau$ is the time-delay between the signals. The angled brackets $\langle . \rangle$ denote the time-average performed over millions of condensate realizations (i.e., the system is ergodic). 

\begin{figure}[t!]
\includegraphics [width=1\columnwidth]{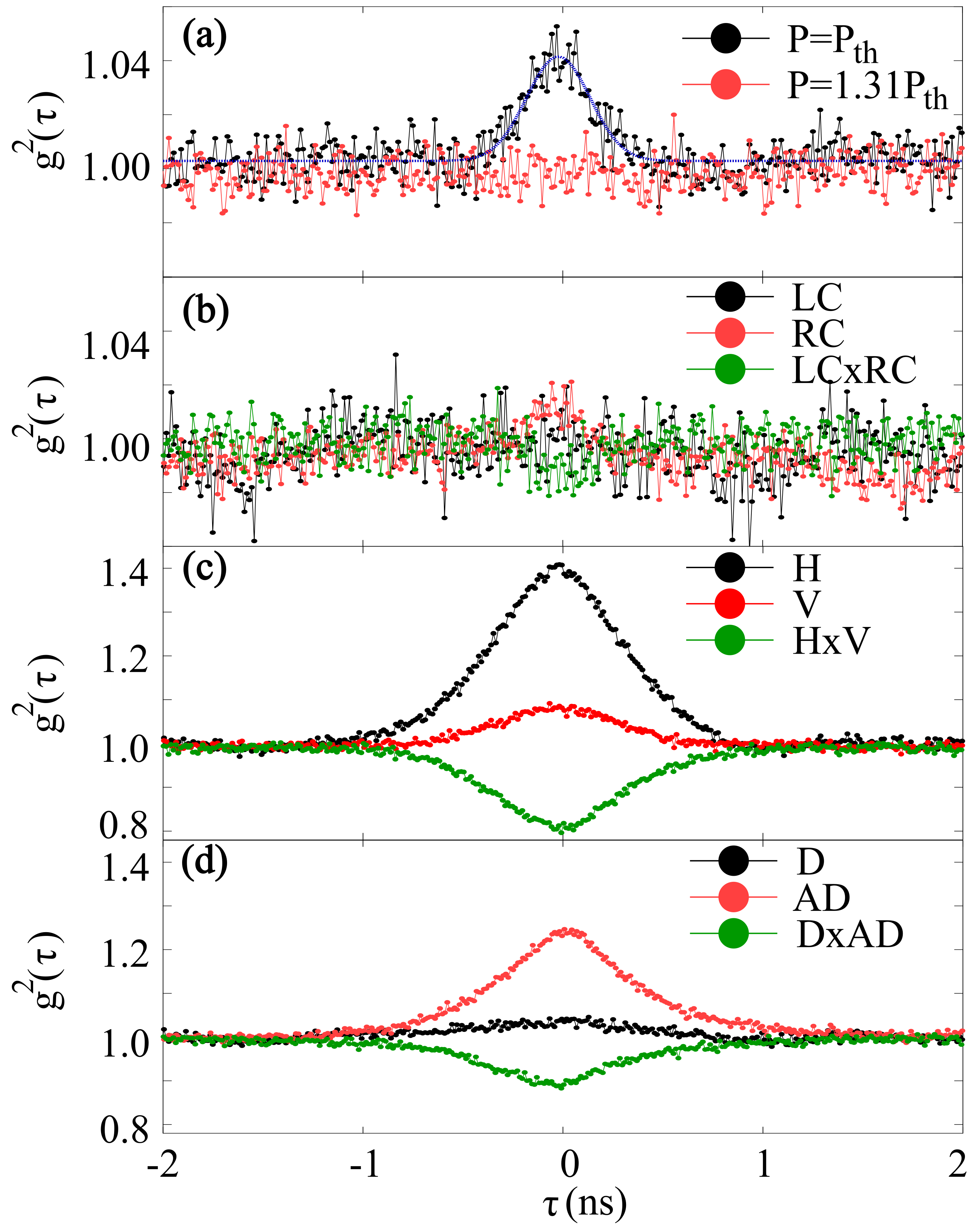}
\centering
\caption{(a) Second order correlation function of the condensate PL, measured without polarization separation at $P=P_{th}$ and $P=1.31P_{th}$. Blue line is a Gaussian fit. (b-d) $g_{i,i}^{(2)}(\tau)$ (black), $g_{j,j}^{(2)}(\tau)$ (red), and $g_{i,j}^{(2)}(\tau)$ (green) at $P=1.31P_{th}$ for LC-RC, H-V, and D-AD polarizations respectfully.}
\label{fig.2}
\end{figure}

Figure~\ref{fig.2}(a) shows the second order correlation function measured with a standard HBT setup (all light enters interferometer) at the condensation threshold ($P = P_{th}$) and above threshold ($P=1.31P_{th}$) for horizontally polarized excitation. One can see slight photon bunching $g^{(2)}(0)\approx 1.04$ at threshold with $\tau_c = 220$~ps second-order coherence time, estimated from a Gaussian function fit. Increasing the excitation density to $P = 1.31 P_{th}$ drives the condensate towards a highly coherent state such that one cannot resolve any signature of photon bunching on top of the shot noise level, $g^{(2)}(\tau)\simeq1$, in agreement with previous studies~\cite{PhysRevLett.100.067402, PhysRevX.6.011026}. We did not observe any measurable deviation from the coherent state for the whole condensate with increasing pump power up to $3.5P_{th}$.
%

Interestingly, while the condensate as a whole becomes highly coherent at $1.31P_{th}$, its $S_{1,2}$ pseudospin projections obey super-Poisson photon distribution as seen from $g_{i,i}^{(2)}(0)>1$ registered for different linear polarization components in Fig.~\ref{fig.2}(c,d). This excess noise can be attributed to the linearly polarized nonresonant optical excitation which results in a condensate pseudospin  approximately forming with random polarization in the equatorial plane of Poincar\'{e} sphere due to the spontaneous breaking of the U(1) symmetry~\cite{Shelykh_PRL2006, Baumberg_PRL2008, PhysRevLett.109.016404} from realization to realization. Moreover, with no stimulated mechanism distinguishing between different linear polarizations of the pseudospin one can intuitively expect it to become susceptible to random walk in the equatorial plane due to polariton-polariton interactions~\cite{Solnyshkov_Semic2007}. This inevitably implies additional noise in the $S_{1,2}$ components equatorial projections. However, any strain-induced birefringence of the microcavity structure will break this planar symmetry~\cite{PhysRevX.5.031002, Kasprzak_PRB2007,KLOPOTOWSKI_SSC2006} as it causes linear polarization splitting in close analogy to an in-plane effective magnetic field applied in the Poincar\'e equatorial plane direction. This field $\boldsymbol{\Omega}_\parallel = (\Omega_x,\Omega_y)^T$ would conventionally lead to pseudospin precession but instead polariton-polariton interactions preferably align the condensate pseudospin parallel with the field, pinning it down~\cite{PhysRevLett.100.067402, PhysRevB.80.195309, Kasprzak_PRB2007, PhysRevB.102.125419}. As this birefringent field comes from the cavity disorder it will randomly vary between sample locations. Moreover, such a field can in general be complex-valued due to the dissipative nature of the system meaning that the split in-plane modes will have different linewidths~\cite{PhysRevX.5.031002}. In accordance with our previous study~\cite{PhysRevB.102.125419}, we will take $\boldsymbol{\Omega}_\parallel$ to be real-valued, causing the pseudospin to become pinned between the diagonal and vertical projections [see Fig.~\ref{fig.3}(b)] corresponding to the dominant polarization in the cavity PL intensity [see Supplemental Information (SI)]. 

Analysis of photon cross-correlations in this {\it pinned} regime gives important insight into the nonlinear and complex dynamics of the pseudospin and its connection with photon statistics. Figures~\ref{fig.2}(c,d) reveal anticorrelated photon fluctuations between orthogonal projections which have identical temporal envelope in their corresponding auto-correlation profiles. This means that fluctuations in one projection inevitably induce fluctuations in the orthogonal one. Namely, the anticorrelated behavior $g_{i,j}^{(2)}(0)<1$ corresponds to temporal switching of the pseudospin direction in the equatorial plane from being parallel to antiparallel with the effective birefringent magnetic field. We observe that this switching has a stochastic nature rather than periodic behavior (see SI). The switching strongly affects the photon statistics of the linearly polarized projections, especially those which are orthogonal to the pinned projection. Indeed, we observe the highest value of $g_{i,i}^{(2)}(0)$ in Fig.~\ref{fig.2}(c,d) for horizontal and anti-diagonal projections. 

Our experiments show significant effect of nonlinear dynamics on the photon statistics which can be controlled through the power of the optical excitation. We remind that the photon auto-correlation $g_{i,i}^{(2)}(0)$ at zero time delay relates directly to a variance ${\sigma_{i}}^2$ and mean value $\bar{n_{i}}$ of the photon distribution according to
\begin{align}\label{eq.3}
g_{i,i}^{(2)}(0)= 1+\frac{{\sigma_{i}}^2-\bar{n_{i}}}{{\bar{n_{i}}}^2}.
\end{align}
With increasing pump power we observe growth of $g_{H,H}^{(2)}(0)$ [see black circles in Fig.~\ref{fig.3}(a)] and signs of excess photon noise with distribution broadening. 
Figure~\ref{fig.3}(a) also shows photon auto-correlations $g_{V,V}^{(2)}(0)$ in red and cross-correlation $g_{H,V}^{(2)}(0)$ in green as a function of pump power. While the horizontal component experiences very strong photon number fluctuations with $g_{H,H}^{(2)}(0)\approx1.9$ above $1.5P_{th}$ the vertical component is in a highly coherent state $g_{V,V}^{(2)}(0)<1.02$. In fact, the large difference in populations between H and V projections and the stochastic switching of the pseudospin, schematically shown in Fig.~\ref{fig.3}(b), leads to a super-Poissonian photon distribution of the horizontal component, coming close to statistics that of thermal light.
\begin{figure}[t]
\includegraphics [width=1\columnwidth]{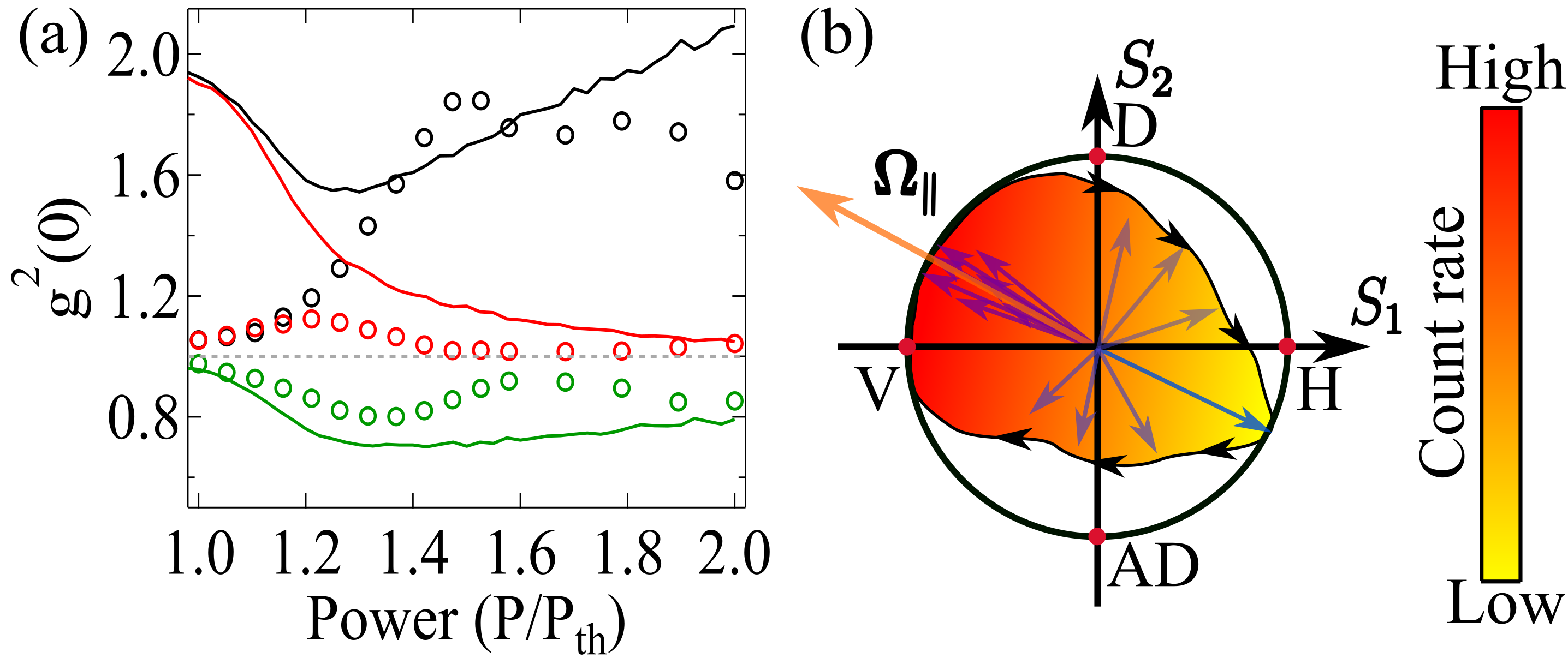}
\centering
\caption{(a) Experimentally measured (circles) and theoretically (solid line) obtained $g_{h,h}^{(2)}(0)$ (black), $g_{v,v}^{(2)}(0)$ (red) and $g_{h,v}^{(2)}(0)$ (green) power dependencies. (b) Schematic representation of the in-plane magnetic field. When pinned, the pseudospin is dominantly orientated parallel $\boldsymbol{\Omega}_\parallel$ (indicated by number of arrows) whereas random {\it switching} events can cause the pseudospin to stabilize briefly anti-parallel, registering as an anticorrelated event between linear polarization components. Color scale represent total correlation photon count rate.}
\label{fig.3}
\end{figure}

Evolution of the pseudospin and correspondent intensity fluctuations are well reproduced by a polariton mean field theory (see SI) as shown in Fig.~\ref{fig.3}(a), except in a region nearby condensation threshold where our time resolution severely limits HBT measurements and standard mean field theories, which rely on large particle numbers in the condensate order parameter, fail to describe high-order correlations~\cite{carmichael2013statistical,   Witthaut_PRA2011}. Nevertheless, numerical simulations using a stochastic generalized Gross-Pitaevskii model in the truncated Wigner approximation~\cite{Wouters_PRB2009, PhysRevB.80.195309} agree with experimental data above $\sim1.3P_{th}$. 

We next investigate how the pseudospin dynamics under elliptically polarized pumping affect the photon statistics. In this case, the exciton reservoir which provides gain to the condensate becomes spin-imbalanced (optically orientated) and follows the laser circular polarization to some degree. The stimulated nature of polariton scattering into the condensate typically preserves the exciton spin and thus usually results in a condensate also co-circularly polarized with the laser~\cite{PhysRevB.91.075305, PhysRevLett.109.036404, PhysRevB.93.205307}. Due to strong anisotropic particle interactions, such spin population imbalance in both the condensate and the reservoir results in a nonlinear effective out-of-plane magnetic field $\boldsymbol{\Omega}_\perp = \Omega_z \hat{\mathbf{z}}$ giving rise to self-induced Larmor precessions ~\cite{Shelykh_PRB2004, Krizhanovskii_PRB2006, Solnyshkov_Semic2007, PhysRevB.80.195309, Ryzhov_PhysRevRes2020,  askitopoulos2020coherence} which are self-sustained periodic orbitals in the dynamical equations of motion. The magnitude of this field can be written $\Omega_z = \alpha S_3 + g (X_+ - X_-)$ where $\alpha$ denotes the polariton-polariton interaction strength, $g$ polariton-exciton interaction strength, and $X_\pm$ are the exciton reservoir spin populations. This imbalance, and the magnitude of $\Omega_z$, can be controlled via ellipticity of the pump laser~\cite{PhysRevB.102.125419}. To adjust the ellipticity ($\epsilon$) of the incident beam we employ a quarter-wave plate (QWP1) placed at the excitation path with an angle $\theta$. The resultant power of left ($P_-$) and right ($P_+$) circularly polarized components of the pump depend on the ellipticity parameter $\epsilon=\sin{(2\theta)}$ following $P_{\pm}=P(1{\mp}\epsilon)/2$, where $P$ is the total pump power. 
\begin{figure}[tb!]
\includegraphics [width=1\columnwidth]{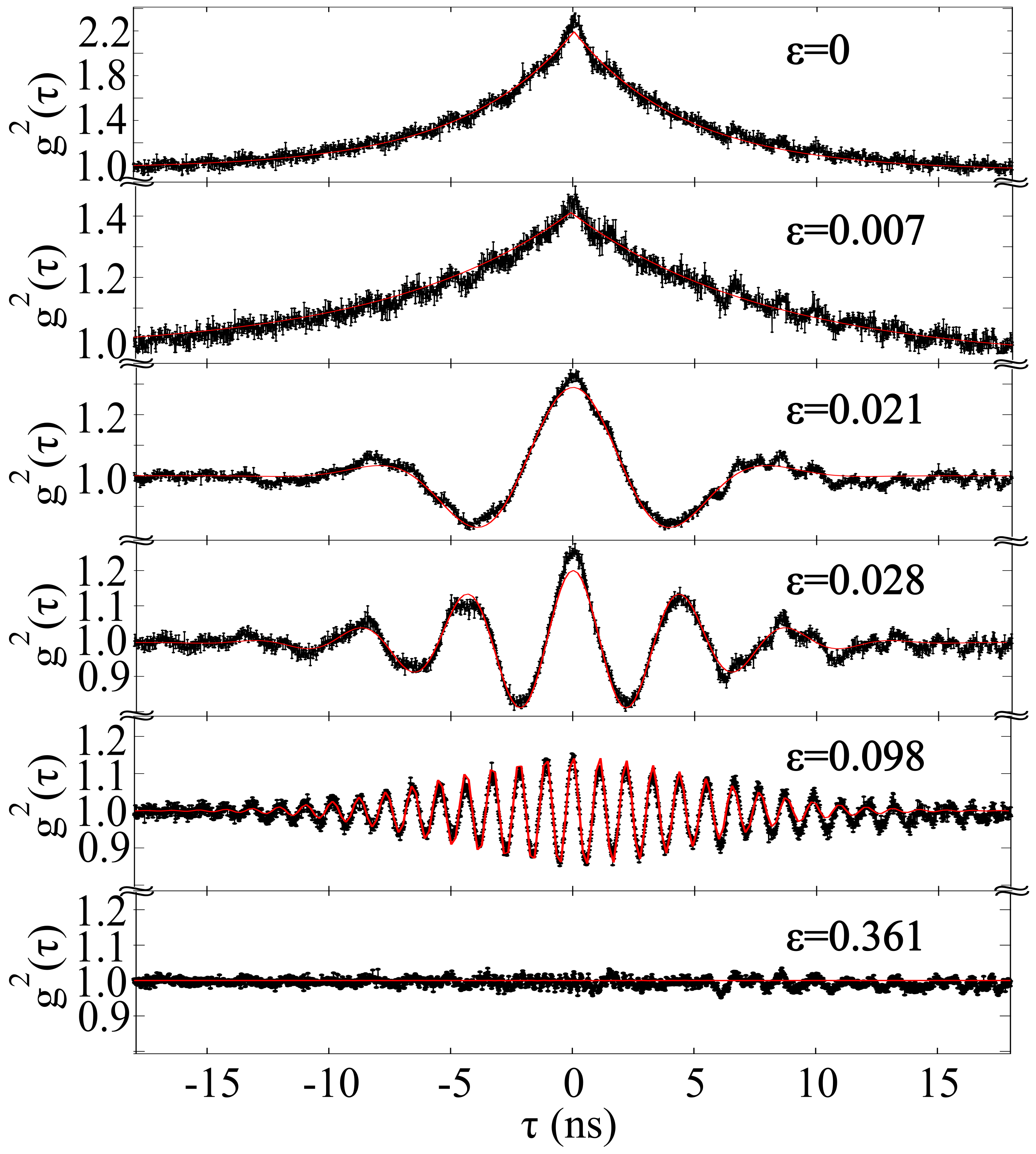}
\centering
\caption{Photon statistics engineering  with excitation polarization ellipticity. $g_{H,H}^{(2)}(\tau)$ at pump power $P=3.5P_{th}$ for range of excitation ellipticities $\epsilon$. Red lines for $\epsilon=0$ and $0.007$ have been fitted by an exponential function $g_{H,H}^{(2)}(\tau)=1+(g_{H,H}^{(2)}(0)-1)e^{-2\mid\tau\mid/\tau_c}$ and the remaining red lines were fitted by a product of a Gaussian function and a cosine function $g_{H,H}^{(2)}(\tau)=1+\cos{(\omega \tau)}(g_{H,H}^{(2)}(0)-1)e^{-\pi(\tau/\tau_c)^2}$ for clarity. $\tau_c$ is correlation time.}
\label{fig.4}
\end{figure}

\begin{figure}[t!]
\includegraphics [width=1\columnwidth]{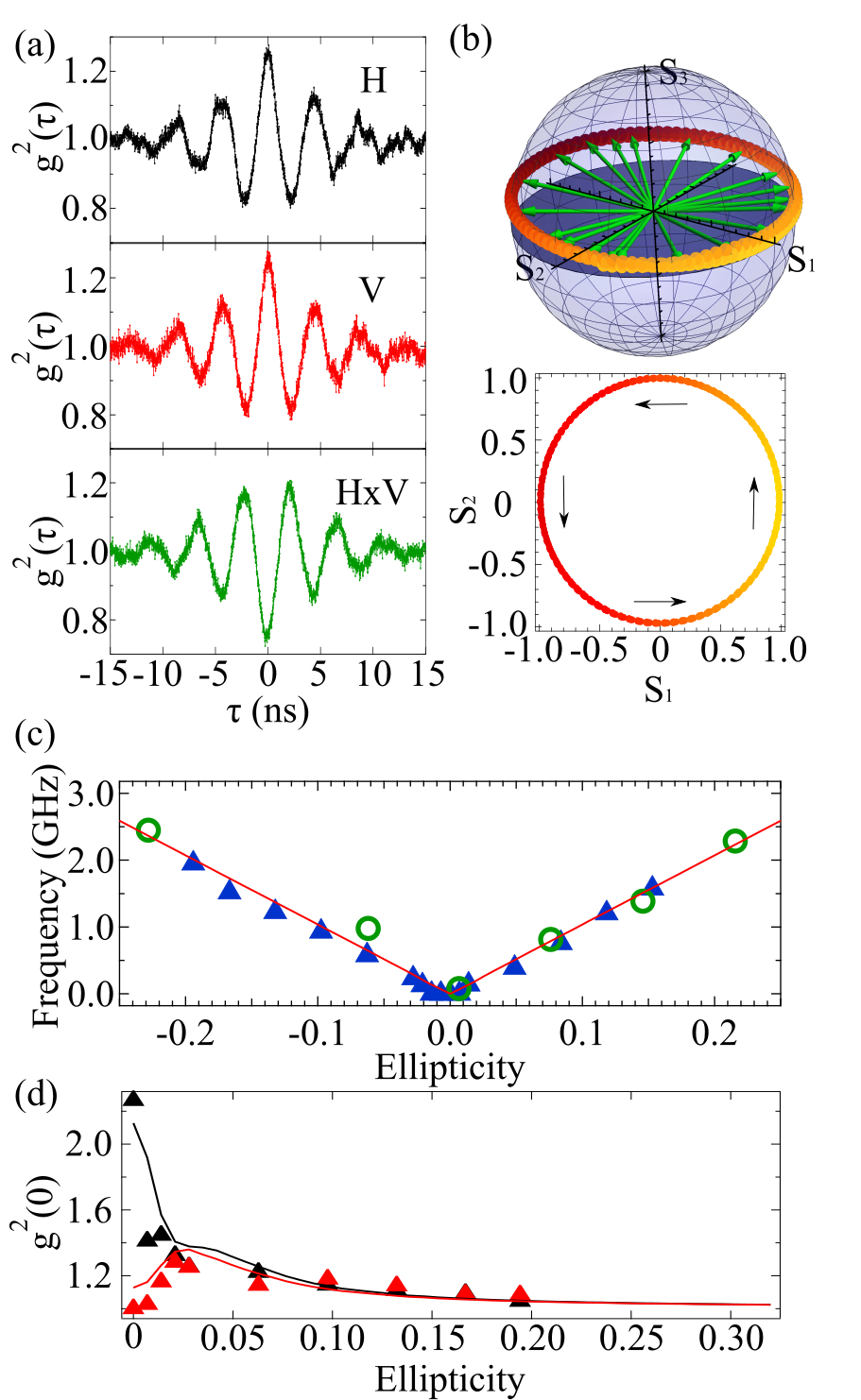}
\centering
\caption{(a) $g_{H,H}^{(2)}(\tau)$ (black), $g^{(2)}_{V,V}(\tau)$ (red), and $g^{(2)}_{H,V}(\tau)$ (green) at $P=3.5P_{th}$ with excitation polarization ellipticity at $\epsilon=0.028$. (b) Calculated normalized pseudospin precession trajectory and its projection onto equatorial plane of Poincar\'{e} sphere. Color scale is for better visualization. Black arrows show direction of trajectory evolution. (c) Measured $g^{(2)}_{i,j}(\tau)$ oscillation frequency (blue triangles) plotted along measured energy splitting (green circles) between LC and RC condensate emission. Red line is splitting between $\psi_+$ and $\psi_-$ from modelling. (d) Measured (triangles) and calculated (solid line) $g_{H,H}^{(2)}(0)$ (black) and $g_{V,V}^{(2)}(0)$ (red) for varying excitation ellipticity.}
\label{fig.5}
\end{figure}

Figure~\ref{fig.4} shows $g_{H,H}^{(2)}(\tau)$ at $P=3.5P_{th}$ for varying ellipticity from 0 (top) to 0.361 (bottom). With linearly polarized excitation ($\epsilon = 0$) horizontal pseudospin projection gives $g_{H,H}^{(2)}(0)~2.4$ with correlation time $\tau_c=9ns$. From $\epsilon = 0.021$ and above an oscillatory behavior in the photon correlations is observed, evidencing nonlinear self-sustained Larmor precessions of the pseudospin due to the combined effects of the nonlinear  field $\boldsymbol{\Omega}_\perp$ and the linear birefringent field $\boldsymbol{\Omega}_\parallel$. The Larmor precession drives harmonic photon number oscillations between the orthogonal components in antiphase which is evidenced by the photon cross-correlations $g_{H,V}^{(2)}$ in the bottom panel of Fig.~\ref{fig.5}(a). To illustrate these dynamics, we simulated the condensate pseudpospin trajectories for the case of $\epsilon = 0.028$. Figure~\ref{fig.5}(b) shows the trajectory of the pseudospin on the Poincar\'e sphere and its projection on the equatorial plane. Assuming that the in-plane birefringent field is weak compared to the out-of-plane field (i.e., $|\boldsymbol{\Omega}_\perp| > |\boldsymbol{\Omega}_\parallel|$) the precession frequency of the pseudospin is approximately dictated by the population imbalance of background excitons and polaritons which is tuned through $\epsilon$~\cite{askitopoulos2020coherence}. In Fig.~\ref{fig.5}(c) we show oscillation frequency extracted from $g_{H,H}^{(2)}(\tau)$ and the experimentally measured energy splitting between the LC and RC polarized cavity emission using a wavemeter. Results from numerical simulations show excellent agreement with the observed trend in splitting. 

While the frequency of the $g^{(2)}_{H,H}(\tau)$ oscillations is increasing linearly with the pump ellipticity their amplitude drops to zero at high ellipticity as the pseudospin becomes predominantly aligned towards circular projections (i.e., $|\mathbf{S}| \approx |S_3|$) converging to a stable fixed point solution close to the poles of the Poincar\'{e} sphere. Thus, the effective out-of-plane magnetic field $\boldsymbol{\Omega}_\perp$ mitigates the role of the in-plane effective field $\boldsymbol{\Omega}_\parallel$ leading to the suppression of photon noise in the horizontal projection. Figure~\ref{fig.4} clearly demonstrates the effect of gradual decrease in $g_{H,H}^{(2)}(0)$ with increasing ellipticity. In Fig.~\ref{fig.5}(d) we plot $g_{H,H}^{(2)}(0)$ and $g_{V,V}^{(2)}(0)$ for increasing ellipticity showing a clear trend of photon noise suppression. Therefore a delicate control over the exciton and polariton spin-imbalance offers a full range of tuneability in photon statistics spanning from super-thermal and thermal distributions at $\epsilon\sim0$ to super-Poissionian distributions and highly coherent states at large $\epsilon>0.2$ in accordance to.

In conclusion, we have realized multi-dimensional photon correlation tomography of an optically trapped spinor polariton condensate across its full polarization basis. Our findings demonstrate how the unique nonlinear mean field dynamics of the condensate pseudospin affect the emitted photon statistics with tunable crossover from super-thermal photon distribution to a coherent state. We identify stochastic linear polarization switching due to the inherent cavity birefringence resulting in polarization sensitive photon bunching, and self-induced Larmor precessions in the GHz frequency range visible as oscillations in the linear polarized photon $g^{(2)}_{i,j}(\tau)$ persistent for more than 10 ns, thus, giving the estimate for a spinor dephasing time. Our findings pave the way towards exploiting nonlinear mean-field dynamics of strongly non-equilibrium bosonic systems to fine control their photon statistics.

The authors acknowledge the support of the UK’s Engineering and Physical Sciences Research Council (grant EP/M025330/1 on Hybrid Polaritonics), the RFBR Project No.20-52-12026 (jointly with DFG), and No.20-02-00919. S.B. acknowledge the support of the RFBR project No.20-31-70001. A.Z. acknowledge financial support from Russian Science Foundation (RScF) Grant No. 20-72-10145.

\newpage

\setcounter{equation}{0}
\setcounter{figure}{0}
\setcounter{section}{0}
\renewcommand{\theequation}{S\arabic{equation}}
\renewcommand{\thefigure}{S\arabic{figure}}
\renewcommand{\thesection}{S\arabic{section}}
\onecolumngrid

\vspace{1cm}
\begin{center}
\Large \textbf{Supplemental Material}
\end{center}

\section{Dynamical mean field equations}
The dynamics of the spinor polariton condensate order parameter is modeled through a set of stochastic driven-dissipative Gross-Pitaevskii equations (Langevin equations) coupled to spin-polarized rate equations describing excitonic reservoirs $X_\pm$ feeding the two spin components $\psi_\pm$ of the condensate. The stochastic part of our model ($\theta_\pm$) was formulated in Refs.~\cite{Read_PRB2009, Wouters_PRB2009} under the so-called truncated Wigner approximation which becomes valid above condensation threshold with large particle numbers in the condensate $\langle n \rangle \gg 1$ such that stimulated effects dominate over spontaneous scattering events. We point out that a more accurate treatment of dissipative many-body quantum systems involves writing a density matrix for the polariton field governed by appropriate master equations~\cite{del2010microcavity}. This approach is beyond the scope of the current study where our modelling concerns the limit of large particle numbers where we show that complex nonlinear mean-field forces have quite dramatic effects on the polariton statistics. The model reads:
\begin{subequations} \label{eq.orig}
\begin{align} 
i & \frac{ d\psi_\sigma}{dt}  =   \frac{1}{2} \Big[ \nu V_\sigma+ i \left( R_1 X_\sigma + R_2 X_{-\sigma} - \Gamma \right) \Big] \psi_\sigma - \nu \frac{\Omega_x}{2} \psi_{-\sigma} + \theta_\sigma(t), \\ 
& \frac{ dX_\sigma}{dt}  =   - \left[ \Gamma_R + R_1 (|\psi_\sigma|^2 + 1) + R_2(|\psi_{-\sigma}|^2 + 1) \right]X_\sigma + \Gamma_s(X_{-\sigma} - X_\sigma) + P_\sigma,\\
& V_\sigma = \alpha_1 |\psi_\sigma|^2  + \alpha_2 |\psi_{-\sigma}|^2 + g_1\left(X_\sigma + \frac{P_\sigma}{W}\right) + g_2\left(X_{-\sigma} + \frac{P_{-\sigma}}{W}\right).
\end{align}
\end{subequations}
Here, $\sigma \in \{+, -\}$ are the two spin indices, $\alpha_{1,2}$ denotes the same-spin (triplet) and opposite-spin (singlet) polariton-polariton interaction strengths and $g_{1,2}$ are the corresponding interactions with the reservoir,  $R_{1,2}$ is the rate of stimulated same-spin and opposite-spin scattering of polaritons into the condensate, and $\Gamma$ is the polariton decay rate, $\Gamma_R$ and $\Gamma_s$ describe the decay rate and spin relaxation of reservoir excitons. In principle, scattering from the reservoirs to the condensate should be dominantly spin-preserving ($R_1$) but in the presence of a (effective) magnetic field ($\Omega_x$) one needs to account for the possibility that particles from the opposite-spin reservoir can scatter ($R_2$) into the condensate~\cite{Solnyshkov_Semic2007, Redondo_NewJouPhys2018}. Some studies work under the approximation that in optical traps the condensate is so well separated from the background reservoir and that blueshift coming from polariton-exciton interactions can be discarded but recent studies~\cite{boozarjmehr2020spatial} have shown that the reservoir is actually not so distant from the condensate and therefore additional polariton condensate blueshift coming from this background reservoir ($g_{1,2}$) should be taken into account. For all results presented we have chosen $\alpha_2 = -0.2 \alpha_1$ and $g_2 = -0.2 g_1$~\cite{Ciuti_PRB1998}. We also include an energy dampening parameter $\nu = 1  -i \nu'$ according to the Landau-Khalatnikov approach~\cite{Read_PRB2009}. Finally, spin-mixing (spin-relaxation) between the reservoirs ($\Gamma_s$) should be taken into account as it can be evidenced as depolarization in the cavity photoluminescence below condensation threshold~\cite{Maialle_PRB1993, Ohadi_PRL2012, Redondo_PRB2019, klaas_nonresonant_2019,pickup2020polariton}. It is naturally quite challenging to understand the full picture of which parameters contribute to different observed effects in experiment and thus we attempt at being as inclusive as possible of different physical mechanisms. 

Although the experiment deals with a birefringent field $\boldsymbol{\Omega}_\parallel = (\Omega_x,\Omega_y)^T$ at a specific angle we will, without any loss of generality, take the splitting to be between horizontal $\psi_H = (\psi_+ + \psi_-)/\sqrt{2}$ and vertical $\psi_V = (\psi_+ - \psi_-)/\sqrt{2}$ polarized modes represented by the real-valued $\Omega_x$ and $\Omega_y=0$. The strength of the white complex noise $\theta_\sigma(t)$ is determined by the scattering rate of polaritons into the condensate
\begin{equation}
\langle \theta_\sigma(t) \theta_{\sigma'}(t') \rangle = 0, \qquad   \langle \theta_\sigma(t) \theta^*_{\sigma'}(t') \rangle  = \frac{R_1 X_\sigma + R_2 X_{-\sigma}}{2} \delta_{\sigma \sigma'} \delta(t-t').
\end{equation}
The active reservoir $X_\sigma$, which feeds the condensate with particles, is driven by a background of high momentum {\it inactive} excitons $P_\sigma$ which do not satisfy energy-momentum conservation rules to scatter into the condensate. Assuming the simplest type of rate equation describing the conversion of optical excitation power into an inactive reservoir in the continuous wave regime we write:
\begin{equation} \label{eq.P}
\frac{dP_\sigma}{dt}  = - (W + \Gamma_I) P_\sigma + \Gamma_s(P_{-\sigma} - P_\sigma) + L_\sigma.
\end{equation}
Here, $W\gg\Gamma_I$ where $W$ is a phenomenological spin-conserving redistribution rate of inactive excitons into active excitons and $\Gamma_I$ is the nonradiative exciton decay rate. Since these inactive excitons also experience spin relaxation $\Gamma_s$ the polarization of $P_\sigma$ will not coincide with that of the incident optical excitation. As the experiment is performed in the continuous-wave regime we can immediately solve the steady state solution of Eq.~\eqref{eq.P} and plug it into Eqs.~\eqref{eq.orig}. For optical excitation parameterized as $\mathbf{L} = L (\cos{(\theta)}, \sin{(\theta)})^T$ where $\theta$ can be understood as a quarter waveplate angle in the experimental setup, we can write the background reservoir as,
\begin{equation}
\begin{pmatrix} P_+ \\ P_- \end{pmatrix} = \frac{L}{W + 2 \Gamma_s} \begin{pmatrix} W \cos^2{(\theta)} + \Gamma_s \\ W \sin^2{(\theta)} + \Gamma_s \end{pmatrix}.
\end{equation}
Here, $L$ is the power of the optical excitation and $\theta$ determines the polarization of the incident light. 

Determining the parameters of Eq.~\eqref{eq.orig} poses a challenge since they will depend in a complicated way on both sample and excitation properties. To overcome this, we implement a random walk algorithm which, in each step, calculates the root-mean-square-error between the experimental and simulation data. The algorithm starts from a random set of parameters (appropriately bounded to remain physical) and repeatedly takes a random step forward in parameter space which is kept if the error is lowered. If the error rises, the step is discarded (go back to previous step) and a new random step is tested. Performing 500 random initializations in the parameter space, with each taking 300 random steps, we determine a set of parameters best fitting the experimental results. The parameters used throughout the manuscript are given in units of $\Gamma$ except of $\nu'$ which is dimensionless: $\Gamma_R = 1.6$; $\Gamma_s = 0.19$; $W = 0.156$; $R_1 = 0.0032$; $R_2 = 0.0027$; $\alpha_1 = 0.00015$; $g_1 = 0.00097$; $\nu' = 0.077$. For Fig.~3(a) and~5(c,d) in the main text we have set $\Omega_x = -0.057$ and $-0.0072$ respectively as they correspond to different sample locations. The theoretical pump threshold value for linearly polarized excitation corresponds to a threshold laser power $L$ defined as $P_{th} = P_\pm(L_{th})$,
\begin{equation}
L_{th} = \frac{2 \Gamma}{ \dfrac{R_1 + R_2}{\Gamma_R} + \nu'(g_1 + g_2[\Gamma_R^{-1} + W^{-1}])}.
\end{equation}

\section{Polarization switching and pinning}

\begin{figure}[h]
\includegraphics [width=1\columnwidth]{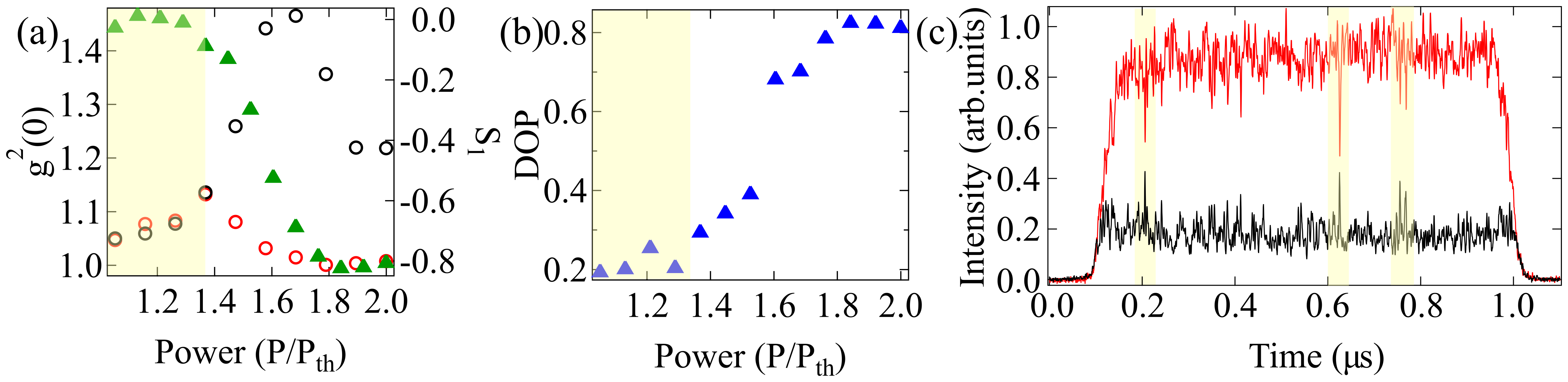}
\centering
\caption{(a) Experimentally measured (circles) $g_{i,j}^{(2)}(0)$ power dependence for H (black) and V (red) polarized light and corresponding integrated $S_1$ component. (b) Integrated measurement of condensate degree of polarization (DOP). Yellow rectangle highlights the pump power range without polarization pinning (low DOP). (c) Measured time series (1$\mu$s excitation pulse) for condensate H (black) and V (red) polarized emission. While pinned to vertical polarization we observe switching of intensity between the polarizations.}
\label{fig.1}
\end{figure}

\end{document}